\documentclass[%
 aip, apl, reprint
 amsmath,amssymb,
 reprint,%
]{revtex4-1}

\usepackage{graphicx}
\graphicspath{{./}{Figure/}}
\usepackage{dcolumn}
\usepackage{bm}

\usepackage{float}
\usepackage[utf8]{inputenc}
\usepackage[T1]{fontenc}
\usepackage{mathptmx}
\usepackage{etoolbox}
\usepackage{siunitx}
\usepackage{textcomp} 
\usepackage{xcolor}
\usepackage[normalem]{ulem}

\usepackage{hyperref} 
\hypersetup{
    colorlinks=true,
    linkcolor=blue, 
    citecolor=blue, 
    urlcolor=black    
}

\makeatletter
\def\@email#1#2{%
 \endgroup
 \patchcmd{\titleblock@produce}
  {\frontmatter@RRAPformat}
  {\frontmatter@RRAPformat{\produce@RRAP{*#1\href{mailto:#2}{#2}}}\frontmatter@RRAPformat}
  {}{}
}%
\makeatother
\begin{document}

\preprint{AIP/123-QED}

\title{Enhanced Polarization Locking in VCSELs}
\author{Zifeng Yuan}
\author{Dewen Zhang}%
\author{Lei Shi}%
\author{Yutong Liu}%
\author{Aaron Danner}
\affiliation{\makebox[\textwidth][l]{National University of Singapore, Department of Electrical and Computer Engineering, Singapore}}

 \email{adanner@nus.edu.sg}
 
\date{\today}

\begin{abstract}

While optical injection locking (OIL) of vertical-cavity surface-emitting lasers (VCSELs) has been widely studied in the past, the polarization dynamics of OIL have received far less attention. Recent studies suggest that polarization locking via OIL could enable novel computational applications such as polarization-encoded Ising computers. However, the inherent polarization preference and limited polarization switchability of VCSELs hinder their use for such purposes.
To address these challenges, we fabricate VCSELs with tailored oxide aperture designs and combine these with bias current tuning to study the overall impact on polarization locking.
Experimental results demonstrate that this approach reduces the required injection power (to as low as \SI{3.6}{\micro\watt}) and expands the locking range. To investigate the impact of the approach, the spin-flip model (SFM) is used to analyze the effects of amplitude anisotropy and bias current on polarization locking, demonstrating strong coherence with experimental results.

\end{abstract}
\maketitle

 Optical injection locking (OIL) is a technique that synchronizes a slave laser (SL) to a master laser (ML) by injecting light from the ML into the SL's cavity, enabling the SL to lase at the same frequency, polarization, and with a fixed phase offset to the ML. OIL has been widely applied in diverse fields such as optical communications~\cite{Kikuchi1984,Chang2003,Liu2014,Kasai2015, GonzalezGuerrero2022, Berisha2024}, optical frequency comb generation~\cite{Temprana2016,Zhou2015,Li2024, Chen2022}, arbitrarily shaped optical pulse generation~\cite{Wu2015,Bhooplapur2017}, and even frequency locking in quantum teleportation~\cite{Paraiso2019, Su2022}. Vertical-cavity surface-emitting lasers (VCSELs) have emerged as widely used semiconductor lasers in OIL due to their compact footprint, low power consumption, and ease of integration into large-scale arrays~\cite{Iga2000,Moser2013,Hirose2014,Danner2006,Heuser2020}.
 
 Research specifically focusing on polarization locking via OIL has received limited attention and has not typically been a primary focus. However, recent studies suggest that polarization locking of VCSELs holds significant potential for novel computational applications, such as polarization-encoded Ising computers and optical neural networks~\cite{Utsunomiya2011, Loke2023, Babaeian2019, Gao2024, Chen2023, Lan2024, Zhang2025, Lim2024, Lim2025, Zhang2025b, Zhang2025Patent, Yuan2025e}, where encoding bits or qubits into orthogonal lasing polarization states presents a promising approach. What many of these applications require is a VCSEL that exhibits equal preference for two orthogonal polarization states~\cite{Utsunomiya2011,Loke2023,Zhang2025}. In other words, an ideal VCSEL for these applications should exhibit easily switchable (unstable) polarization states, which are better suited for achieving polarization locking. However, typically VCSELs inherently exhibit gain anisotropy caused by strain or crystal anisotropy, resulting in a polarization state with predominantly higher gain~\cite{Gehrsitz2000}. This characteristic presents challenges for polarization locking: for VCSELs with strong anisotropy, the polarization state becomes fixed and resistant to switching, or achieving polarization locking requires a high injection power~\cite{Loke2023, Zhang2025}.

Various approaches have been explored to achieve polarization control and to investigate polarization switching or its suppression in VCSELs~\cite{ChangHasnain1991,  Choquette1994, Verschaffelt2001, Martin1997, JansenVanDoorn1996, Panajotov2000, Panajotov2001, Choquette1994b, Panajotov2003, Tan2012, Hallstein1997, Iba2011, Ostermann2005, Huang2007, ChangHasnain2012, Kopp2003, Hodgkinson2002, Kopp1998, Jia2023, Xie2020, Wen2021, Li2018, Seghilani2016, Yuan2025b, Yuan2025c, Yuan2025d}. Among these, several methods have been introduced to control anisotropic gain~\cite{ChangHasnain1991, Choquette1994, JansenVanDoorn1996, Panajotov2000, Panajotov2001, Choquette1994b, Panajotov2003}, including tailoring the aperture shape~\cite{Choquette1994b, Panajotov2003}. Other approaches involve anisotropic current injection mechanisms~\cite{Tan2012} and optical injection locking~\cite{Hallstein1997, Iba2011}. Additionally, polarization control has been demonstrated using surface gratings~\cite{Ostermann2005, Huang2007, ChangHasnain2012}, liquid crystals~\cite{Kopp2003, Hodgkinson2002, Kopp1998}, and metasurface integration~\cite{Jia2023, Xie2020, Wen2021, Li2018, Seghilani2016}. Many of these studies have shown that polarization switching occurs when gain anisotropy or amplitude anisotropy is reduced, leading to polarization instability in VCSELs.

To enhance the polarization locking performance of VCSELs, this work builds upon prior studies~\cite{Choquette1994b, Panajotov2003} that leverage aperture engineering to control polarization properties. We fabricate VCSELs with tailored aperture structures to control amplitude anisotropy, specifically VCSELs which yield less anisotropy in polarization than is typical, thereby improving polarization switchability. Furthermore, we evaluate the OIL performance of these VCSELs under different bias currents. The paper is structured as follows: first, we detail the experimental setup and methods. This is followed by experimental results and comparison with the state-of-the-art in OIL, supported by spin-flip model (SFM) simulations, which illustrate how amplitude anisotropy and bias current influence polarization locking.

\begin{figure*}
\centering 
\includegraphics[height=0.4\textheight]{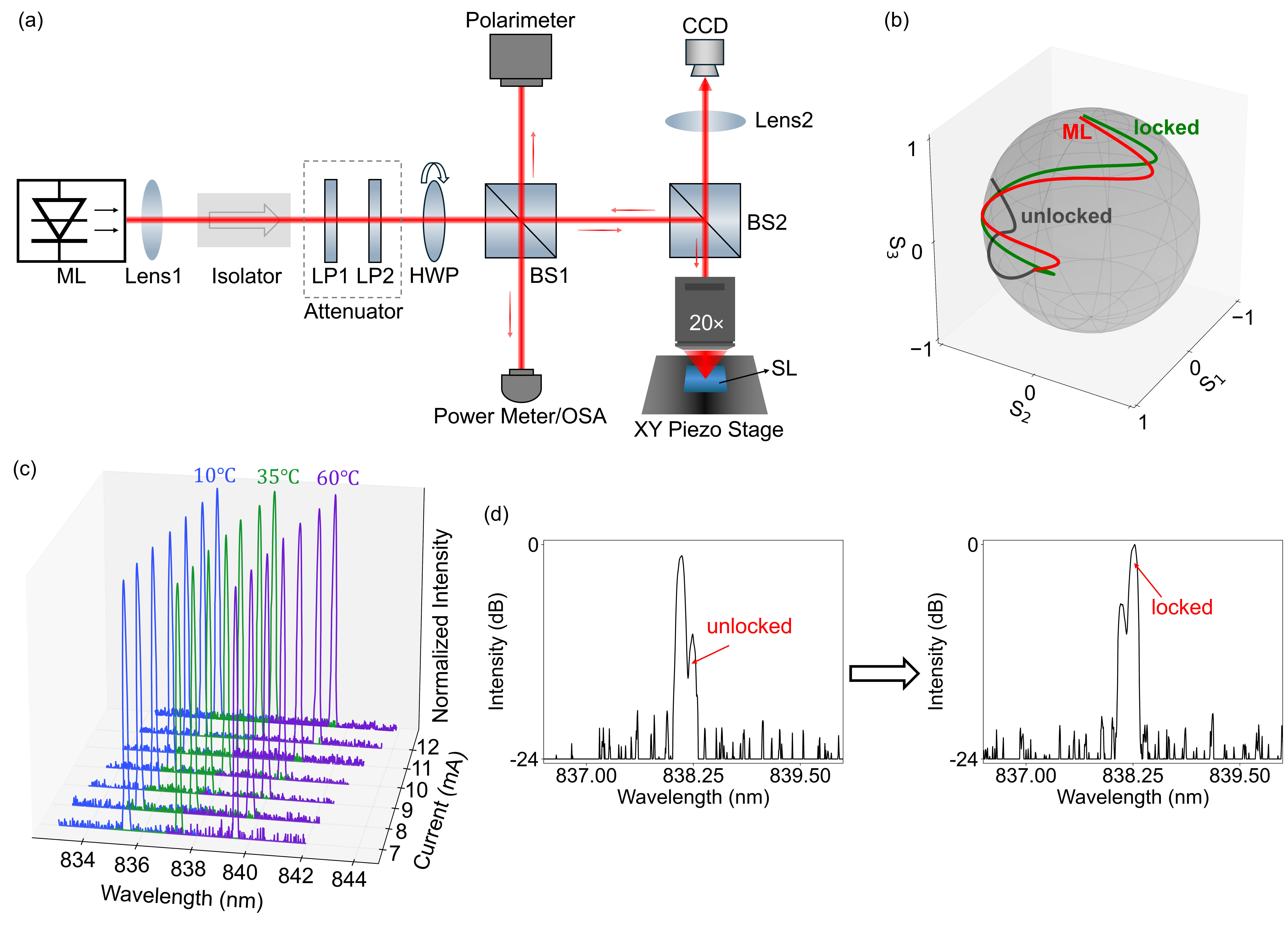} 
\caption{(a) Experimental schematic illustrating the free-space OIL setup. ML: master laser, SL: slave laser, LP: linear polarizer, HWP: half-wave plate, BS: beam splitter, OSA: optical spectrum analyzer. (b) Poincaré sphere showing the polarization states of the ML, the SL unlocked and locked. (c) Spectral of the ML under varying operating conditions. (d) Spectral of the SL in the unlocked and locked states. The peak intensity has been shifted to 0 dB.}
\end{figure*}

The configuration of our experimental setup is shown in Fig. 1(a). It begins with collimated light from the VCSEL ML, emitting light with a center wavelength tunable from 835.57 nm to 841.47 nm, controlled by a temperature controller and a laser diode controller. The spectrum of the emitted light is illustrated in Fig. 1(c). The injection power is modulated by rotating two linear polarizers (LP1 and LP2), while a half-wave plate (HWP) or quarter-wave plate (QWP) is used to adjust the polarization state of the ML light. The light is then directed through a 50:50 beam splitter (BS1),  with half sent to a port that employs a power meter for monitoring the injection power, or a polarimeter (Thorlabs, PAX1000IR1) for measuring the state of polarization of collimated input light using the rotating-wave-plate method, providing azimuth and ellipticity, or an optical spectrum analyzer (OSA) for measuring the spectrum of the ML light. The other half of the beam continues through another 50:50 beam splitter (BS2) and is directed into the fabricated VCSELs on the chip, serving as the SLs. Accurate focusing of the ML light onto the SL’s cavity is achieved using a 20×NIR MO and an XY piezo stage. Current is supplied to the SL chip via probing. The output light from the SL passes through BS2 and BS1, ultimately reaching a port used to measure the output power, polarization state, or spectrum of the SL light. To ensure that parasitic light from the ML does not affect the polarimeter measurements, we first turn off the SL while keeping the ML on. The optical power detected at the polarimeter port is measured and found to be negligible.

An example of polarization locking outcomes is illustrated in Fig. 1(b) and Fig. 1(d), based on the results from the setup shown in Fig. 1(a). Some details such as bias current, frequency detuning and injection power, are further explored in this study. As shown in Fig. 1(b), the SL’s polarization state aligns with the ML under sufficient injection power (locked condition), while it deviates from the ML when the injection power is insufficient (unlocked condition). Fig. 1(d) depicts the SL spectrum under locked and unlocked conditions, with both panels sharing the same normalized power scale. In the left panel (unlocked), with no ML injection, the dominant peak corresponds to the SL’s natural polarization state. The goal is to lock the second-highest peak, which has an orthogonal polarization state. To achieve this, the ML’s polarization is adjusted to be orthogonal to the SL’s natural state, and its wavelength is tuned to align with the second-highest peak. In the right panel (locked), sufficient injection power from the ML causes this target peak to become dominant, while the first peak is suppressed.

\begin{figure*}
\centering
\includegraphics[height=0.95\textheight]{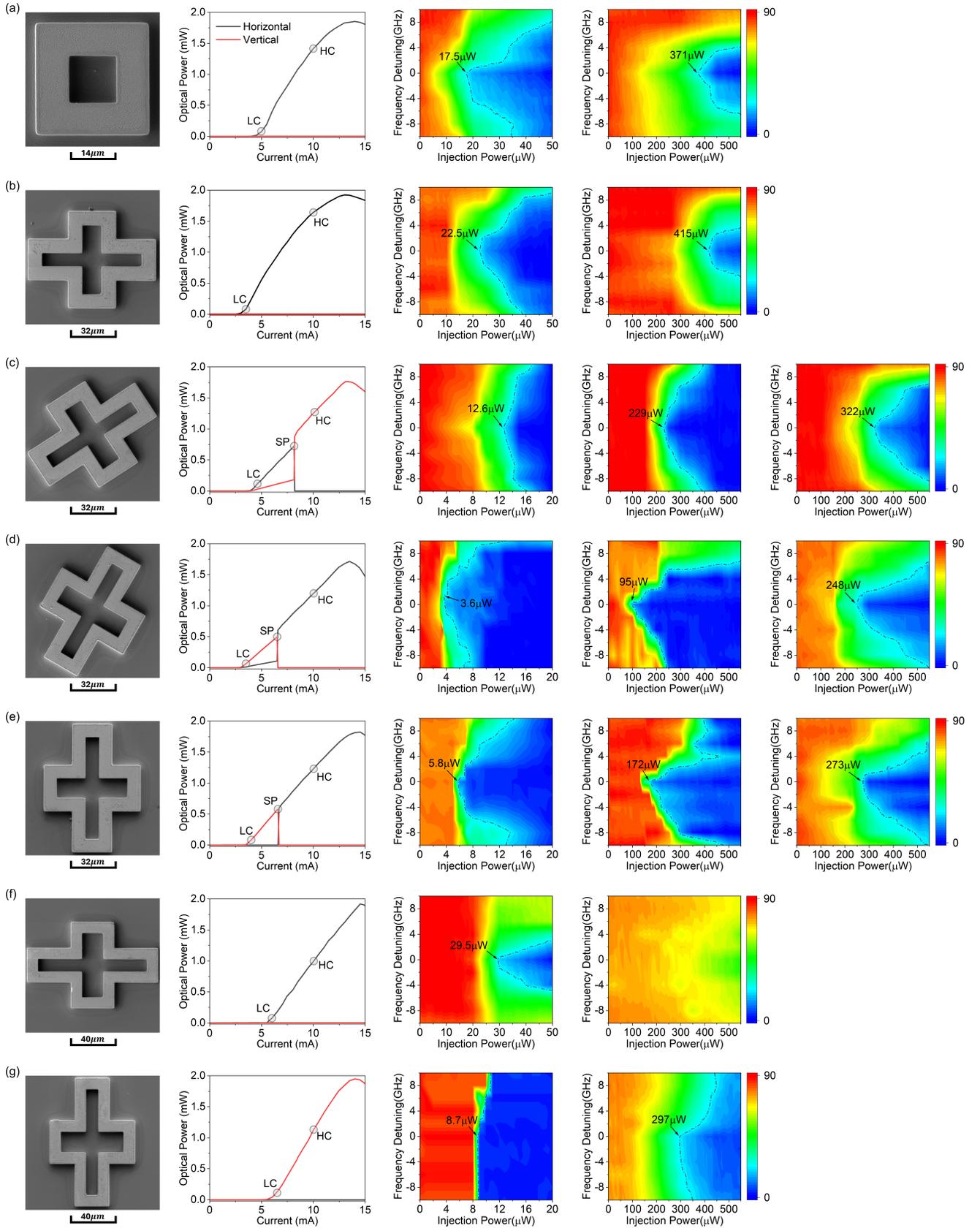}
\caption{OIL performance for VCSELs with tailored aperture designs. (a)–(g) correspond to: (a) square aperture; (b)–(e) cruciform aperture (1:0.7 aspect ratio, rotated by 0\textdegree{}, 30\textdegree{}, 60\textdegree{}, and 90\textdegree{}); (f)–(g) cruciform aperture (1:0.6 aspect ratio, without and with 90\textdegree{} rotation). The first column shows scanning electron microscope images, the second column presents LI curves for horizontal and vertical polarizations, and the remaining columns display locking diagrams at low current (LC), switching point (SP, omitted for non-switchable VCSELs), and high current (HC). The locking diagrams illustrate the absolute polarization difference, $\Delta \theta = |\theta_{\text{SL}} - \theta_{\text{ML}}|$, where $\theta_{\text{SL}}$ and $\theta_{\text{ML}}$ are the absolute azimuths of SL and ML measured by the polarimeter. A dashed line indicates the boundary between unlocked and locked states. The minimum injection power required for successful polarization locking is highlighted.}
\end{figure*}

\begin{table*}
\caption{Summary of OIL performance from some recent papers (2000 to present).}
\begin{ruledtabular}
\begin{tabular}{ccccccc}
Year & Minimum injection power & Locking range  & Wavelength (nm) & SL Laser type & Ref. \\ \hline

$\cdots$& $3.6\,\mu\text{W}$ & $\Delta f = (-10,+10)\,\text{GHz}$  & 838.25 & VCSEL & This work \\
2024 & $2.5\,\mu\text{W}$ & $\Delta f = -12.46\,\text{GHz}$$^{*}$  & 850 & VCSEL & \cite{Zhang2025}\\
2023 & $94\,\mu\text{W}$ & $\Delta f = -23.58\,\text{GHz}$$^{*}$ &  976.77 & VCSEL & \cite{Pfluger2023}\\
2023 & $-19\,\text{dB}$ & $\Delta f = (0,+30)\,\text{GHz}$ &  1547 & VCSEL & \cite{Yokota2023} \\
2017 & $25.1\,\mu\text{W}^{*}$ & $\Delta f = +0.5\,\text{GHz}^{*}$ &1540 & VCSEL & \cite{Bhooplapur2017} \\
2017 & $11\,\mu\text{W}^{*} $ & $\cdots$ & 850 & VCSEL & \cite{Nazhan2017} \\
2017 & $\sim1\,\mu\text{W}^{*} $ & $\Delta f = (-8,-2),\,(0,+2)\,\text{GHz}$$^{*}$ & 1540.91 & VCSEL & \cite{DenisleCoarer2017} \\
2017 & $3.16\,\mu\text{W}$ & $\cdots$ & 1550 & VCSEL & \cite{Jignesh2017} \\
2017 & $10\,\text{dB}^{*}$ & $\Delta f = -1.01\,\text{GHz}^{*}$ & 1542.4 & VCSEL & \cite{Xiao2017}\\
2016 & $0.8\,\text{mW}$$^{*}$ & $\Delta f = -12.46\,\text{GHz}^{*}$ & 850 & VCSEL & \cite{Lu2016b} \\
2016 & $2\,\text{mW}$ & $\Delta f = (-19.37,-12.91)\,\text{GHz}$$^{*}$  & 681.72–682.12 & VCSEL & \cite{Lu2016a} \\
2016 & $10\,\text{mW}$$^{*}$ &  $\Delta f = -63.06\,\text{GHz}^{*}$  & 1542.35 & VCSEL & \cite{Fontaine2016} \\
2016 & $17.23\,\text{mW}^{*}$ & $\cdots$  & 1537.95 & VCSEL & \cite{Prior2016} \\
2014 & $50\,\mu\text{W}$ & $\Delta f = (-2.45,+1.55)\,\text{GHz}$  & 1538.01–1538.24 & VCSEL & \cite{Lin2014} \\
2013 & $\cdots$ & $\Delta f = (-20,+15)\,\text{GHz}$  & 1550 & VCSEL & \cite{AlSeyab2013}\\
2011 & $26.9\,\mu\text{W}$ & $\Delta f = (-12,+8)\,\text{GHz}$ & 1550 & VCSEL & \cite{Perez2011} \\ 
2011 & $1\,\text{mW}$ & $\Delta f = (-10, 0)\,\text{GHz}$ & 850 & VCSEL & \cite{Qader2011}\\ 
2011 & $\cdots$ & $\Delta f = (-9,+2)\,\text{GHz}$ & 1550 & VCSEL & \cite{AlSeyab2011} \\ 
2009 & $3.981\,\text{mW}$ & $\Delta f = (-15.9,+26.98)\,\text{GHz}$$^{*}$ & 1300 & VCSEL & \cite{Hayat2009} \\
2008 & $7.9\,\mu\text{W}$ & $\Delta f=(-12.64,+12.64)\,\text{GHz}$$^{*}$ &  1540.3 & VCSEL & \cite{Jeong2008} \\
2008 & $<63\,\text{mW}$$^{*}$ & $\Delta f=(-47,+67)\,\text{GHz}$ &  1550 & VCSEL & \cite{Lau2008} \\
2007 & $<25\,\mu\text{W}^{*}$ & $\cdots$ & 1550 & VCSEL & \cite{Zhao2007} \\
2006 & $13.2\,\mu\text{W}$ & $\Delta f=(-84,-4),(+50,+76)\,\text{GHz}$$^{*}$  & 850 & VCSEL & \cite{Gatare2006} \\
2003 & $30\,\mu\text{W}$ & $\Delta f=(-18.73,-2.5)\,\text{GHz}$$^{*}$ & 1550 & VCSEL & \cite{Chang2003}\\
\hline 
\\[-0.2em] 
2024 & $1\,\mu\text{W}$ & $\Delta f = (-1.799 ,+1.195)\,\text{GHz}$  & 1549.315 & Edge(DFB) & \cite{Li2024} \\
2024 & $3.16\,\mu\text{W}$ & $\Delta f = +9.40\,\text{GHz}$  & 1552.6 & Edge(DFB) & \cite{Chen2024} \\
2024 & $20\,\text{nW}$ & $\Delta f = +0.62\,\text{GHz}$  & 1550 & Edge(DFB) & \cite{Wen2024} \\
2024 & $1\,\mu\text{W}$$^{*}$ & $\Delta f = (0 ,+6)\,\text{GHz}$$^{*}$  & 1575.4 & Edge(DFB) & \cite{Luo2024} \\
2024 & $1\,\text{mW}$$^{*}$ & $\Delta f = -50\,\text{GHz}$$^{*}$  & 1550 & Edge(DFB) & \cite{Liu2024} \\
2024 & $-12\,\text{dB}$ & $\Delta f = (-15, +5)\,\text{GHz}$  & $\cdots$ & Edge(DFB) & \cite{Berisha2024} \\
2022 & $-45\,\text{dB}$$^{*}$ & $\Delta f = (0,+4.55)\,\text{GHz}$$^{*}$ &  1550 & Edge(DFB) & \cite{Herrera2022} \\
2022 & $\cdots$ & $\Delta f = (+12.07,+25.37)\,\text{GHz}$ &  1556 & Edge(DFB) & \cite{Lihachev2022} \\
2022 & $1\,\mu\text{W}$ & $\Delta f = +35\,\text{MHz}$ &  1550 & Edge(DFB) & \cite{Su2022} \\
2018 & $0.316\,\text{nW}$ & $\Delta f = +10\,\text{GHz}$ & 1550 & Edge(DFB) & \cite{Kakarla2018} \\
2015 & $1\,\mu\text{W}$ & $\Delta f = (-0.5,+0.5)\,\text{GHz}$ &  1550 & Edge(DFB) & \cite{Zhou2015}\\
2011 & $1\,\text{mW}$ & $\Delta f = (-0.5,+0.5)\,\text{GHz}$$^{*}$ & 1566 & Edge(DFB) & \cite{Fice2011} \\
\hline 
\\[-0.2em] 
2022 & $-15\,\text{dB}$ & $\cdots$ &  1302 & Edge(FP) & \cite{Chen2022} \\
2018 & $2\,\text{mW}$ & $\Delta f = +10\,\text{GHz}$ & 1556 & Edge(FP) & \cite{Liu2018}\\
2016 & $300\,\mu\text{W}$ & $\cdots$  & 399 & Edge(FP) & \cite{Saxberg2016} \\
2016 & $500\,\mu\text{W}$ & $\Delta f = (-2.5,+2.5)\,\text{GHz}$  & 461 & Edge(FP) & \cite{Pagett2016}\\
2012 & $63.1\,\mu\text{W}$ & $\Delta f = (-36.8,+36.8)\,\text{GHz}$$^{*}$ & 1563.22 & Edge(FP) & \cite{Chi2012} \\
\hline 
\\[-0.2em] 
2022 & $-27\,\text{dB}$ & $\Delta f = (-0.12,+0.12)\,\text{GHz}$ &  1550 & \text{InP-Si$_3$N$_4$} lasers & \cite{GonzalezGuerrero2022} \\

2008 & $28\,\text{mW}$ & $\cdots$  & 1550 & Erbium-Doped Fiber Laser & \cite{Chen2008} \\
2006 & $\cdots$ & $\Delta f=(-7,+7)\,\text{MHz}$  & 1064 & Nd:YVO4 Laser & \cite{Kurtz2005} \\

\end{tabular}
\end{ruledtabular}
\begin{flushleft}
\footnotesize{Edge (FP): Fabry-Pérot edge-emitting lasers; Edge (DFB): distributed-feedback edge-emitting lasers.}

\footnotesize{$^{*}$ Calculated or estimated using other information provided in the respective papers.} \\
\end{flushleft}
\end{table*}

In this study, our objective is to enhance the OIL performance, aiming to reduce the injection power required for polarization locking and to expand the locking range. First, we fabricate seven aperture designs, corresponding to Fig. 2(a)–(g) in the following order: square, cruciform (1:0.7 aspect ratio), cruciform (1:0.7 aspect ratio) rotated by 30\textdegree{}, 60\textdegree{}, and 90\textdegree{}, and cruciform (1:0.6 aspect ratio) without and with 90\textdegree{} rotation. The epitaxial wafer used to fabricate VCSELs consists of a GaAs (100) substrate, 40 pairs of bottom Si:GaAs distributed Bragg reflectors (DBRs), GaAs quantum wells, and 28 pairs of top C:GaAs DBRs, with a high-Al oxide layer embedded within the top DBRs. The rotation angle is defined as the angle between the longer arm of the cruciform aperture and the primary flat of a (100) GaAs wafer, which we define as the horizontal direction.
First, a backside n-type contact (AuGe/Ni/Au) is deposited via electron beam evaporation. The top p-type contacts are then defined using photolithography, followed by Ti/Pt/Au evaporation and lift-off.
To form mesas, a SiO$_2$ hard mask is deposited via plasma-enhanced chemical vapor deposition (PECVD), followed by photolithography and subsequent etching of unprotected SiO$_2$ with buffered oxide etch (BOE). The mesa structures are then formed using dry etching via inductively coupled plasma (ICP). After mesa etching, The high-Al layer undergoes wet oxidation at 450$^\circ$C, with the oxidation rate precisely controlled at 0.1 $\mu$m/min by adjusting the N$_2$ flow rate and water steam temperature.

These tailored aperture structures influence the amplitude anisotropy along two orthogonal directions (horizontal and vertical), which in turn determines the polarization state and its switching properties under increasing bias currents. The corresponding switching properties are illustrated in the light power output versus current (LI) curves, measured at room temperature with a current bias ranging from 0 to 15 mA, as shown in Fig. 2. Here, "Horizontal" refers to the dominant polarization direction in square-aperture VCSELs, which is naturally aligned with the primary flat of a (100) GaAs wafer. The polarimeter is calibrated accordingly to ensure that measured polarization states are consistently referenced across all devices. "Vertical" corresponds to the orthogonal direction, and these definitions serve as fixed reference points for all polarization measurements.

To illustrate the impact of aperture engineering on OIL, we focus on the role of built-in gain anisotropy on the polarization state. On our wafer, the VCSELs inherently exhibit higher gain anisotropy in the horizontal polarization state. For example, in Fig. 2(a), square apertures are geometrically symmetric along the horizontal and vertical orthogonal directions. However, due to the inherent gain anisotropy, the horizontal polarization state becomes dominant, and no polarization switching occurs as the bias current increases, as shown in the LI curve in Fig. 2(a). By modifying the aperture shapes, such as using cruciform designs, we can either strengthen (e.g., Fig. 2(b), with a ratio of 1:0.7 and a longer horizontal arm) or mitigate (e.g., Fig. 2(e), with a ratio of 1:0.7 rotated by 90\textdegree{}, where the vertical arm is longer) the built-in amplitude anisotropy. For instance, in cruciform (ratio 1:0.7) rotated by 90\textdegree{}, the natural horizontal anisotropy is suppressed, resulting in a less stable polarization state that is easier to switch with increasing injection current, as shown in the LI curve in Fig. 2(e). Conversely, greater anisotropy leads to more stable polarization states that are harder to switch, as observed in the LI curve in Fig. 2(b).

Next, we bias the current at three distinct levels to evaluate its impact on OIL performance of each VCSEL: low current (LC), just above the threshold current; the switching point (SP), where polarization switching occurs for part of the fabricated VCSELs; and high current (HC), fixed at 10~mA for all VCSELs. The corresponding OIL performance is presented in the locking maps on the right in Fig 2. From left to right, the maps sequentially present the performance for LC, SP (omitted for unswitchable VCSELs), and HC.

We evaluate the OIL performance of all the fabricated VCSELs in our optical setup by varying the injection power and frequency detuning between the ML and the SL within the range of $\pm$10~GHz. In this setup, the polarization state of the ML is adjusted to be orthogonal to the SL. The results show that VCSELs exhibiting polarization switching properties (Fig.~2(c)--(e)) have better OIL performance. At the same bias current (10~mA at HC), the minimum injection power required for these switchable VCSELs is 248~$\mu$W, as shown in Fig.~2(d). In contrast, non-switchable VCSELs, such as Fig.~2(a), require a higher minimum injection power of 371~$\mu$W. Additionally, the locking range (indicated by the blue area) is broader for switchable VCSELs compared to non-switchable ones. By lowering the bias current from HC to LC, the minimum injection power can be further reduced from 248~$\mu$W to 3.6~$\mu$W (Fig.~2(d)).

This work achieves a relatively low minimum injection power when compared to other reports of injection-locked VCSELs. Although not absolutely the lowest, the method provides a complementary approach to further enhance OIL performance limits. A detailed comparison of OIL performance of various lasers in recent years, including VCSELs, Fabry-Pérot edge-emitting lasers (Edge(FP)), distributed-feedback edge-emitting lasers (Edge(DFB)), and others, is presented in Table~1. Among these, Edge(DFB) lasers achieve the lowest minimum injection power; however, their locking range is narrower compared to VCSELs.  Our method simultaneously achieves low injection power over a wide locking range.

To investigate, from a theoretical standpoint, how the proposed method influence VCSEL OIL performance, we employ the SFM~\cite{Martin1997, AlSeyab2011, AlSeyab2013}, which considers spin sublevels of the conduction and valence bands in quantum well lasers. Spin-up ($n^+$) and spin-down ($n^-$) electron recombination generate two lasing transitions: right and left circularly polarized states ($\tilde{E}_+$ and $\tilde{E}_-$). These transitions are coupled through the total carrier density $N = (n^+ + n^-)/2$ and the spin carrier density difference $m = (n^+ - n^-)/2$. These field components can be rewritten in terms of $x$- and $y$-polarization states. The rate equations for an optically injected VCSEL are expressed as follows:

\vspace{-4mm}

\begin{eqnarray}
\frac{dE_x}{dt} &=& \kappa \left[(N-1)E_x - mE_y (\sin{\Delta \phi} + \alpha \cos{\Delta \phi})\right] \nonumber \\
&& - \gamma_a E_x + K_{\text{inj}} E_{\text{injx}} \cos{\Delta_x}, \\
\frac{dE_y}{dt} &=& \kappa \left[(N-1)E_y + mE_x (\alpha \cos{\Delta \phi} - \sin{\Delta \phi})\right] \nonumber \\
&& + \gamma_a E_y + K_{\text{inj}} E_{\text{injy}} \cos{\Delta_y}, \\
\frac{d\phi_x}{dt} &=& \kappa \left[\alpha(N-1) + m \frac{E_y}{E_x} (\cos{\Delta \phi} - \alpha \sin{\Delta \phi})\right] \nonumber \\
&& - \Delta \omega - \alpha \gamma_a + K_{\text{inj}} \frac{E_{\text{injx}}}{E_x} \sin{\Delta_x}, \\
\frac{d\phi_y}{dt} &=& \kappa \left[\alpha(N-1) - m \frac{E_x}{E_y} (\alpha \sin{\Delta \phi} + \cos{\Delta \phi})\right] \nonumber \\
&& - \Delta \omega + \alpha \gamma_a + K_{\text{inj}} \frac{E_{\text{injy}}}{E_y} \sin{\Delta_y},\\
\frac{dN}{dt} &=& -\gamma \left[N\left(1 + E_x^2 + E_y^2\right) - \eta - 2mE_yE_x \sin{\Delta \phi}\right], \\
\frac{dm}{dt} &=& -\gamma_s m - \gamma \left[m\left(E_x^2 + E_y^2\right)\right] + 2\gamma N E_yE_x \sin{\Delta \phi}.
\end{eqnarray}

\begin{table}[h!]
\caption{SFM parameters.}
\centering
\begin{ruledtabular}
\begin{tabular}{@{\hskip 0.65in}c@{\hskip 0.5in}r@{\hskip -0.2in}l@{\hskip 0.75in}}
\textbf{Symbol} & \textbf{Value} & \\
\hline
$\gamma_a$ & $0$ / $-30$ & $\text{GHz}$ \\
$\gamma_s$ & $50$ & $\text{GHz}$ \\
$\gamma_p$ & $2$ & $\text{GHz}$ \\
$\gamma$ & $1$ & $\text{GHz}$ \\
$\eta$ & $[1.1, 3.4]$ & \\
$\kappa$ & $300$ & $\text{GHz}$ \\
$\alpha$ & $3$ & \\
$K_{\text{inj}}$ & $35.5$ & $\text{GHz}$ \\
\end{tabular}
\end{ruledtabular}
\end{table}

\begin{figure*} 
\centering 
\includegraphics[height=0.33\textheight]{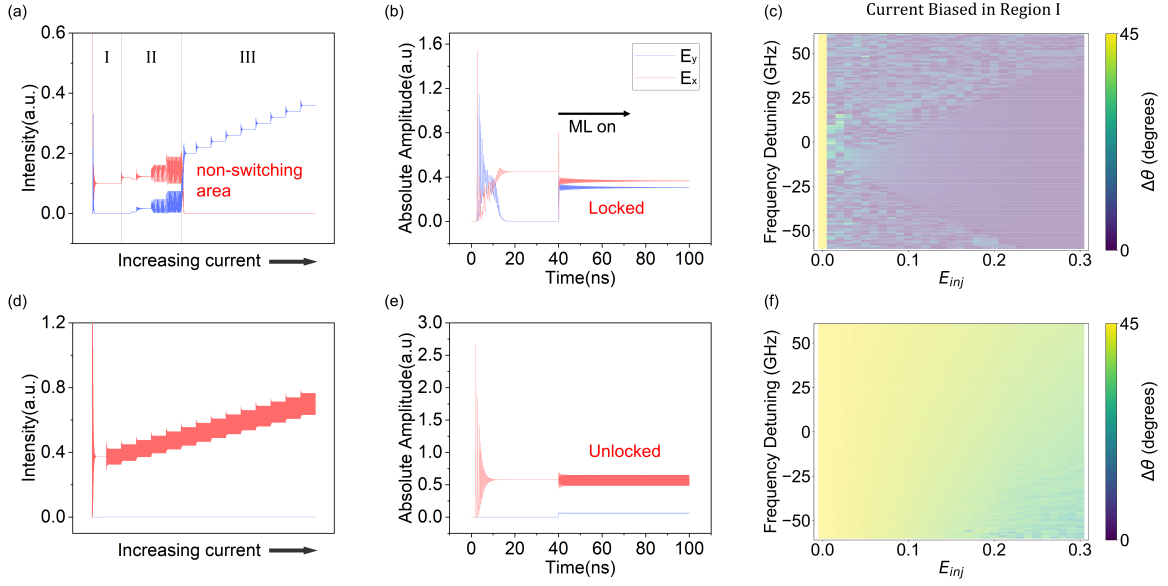} 
\caption{
VCSEL polarization dynamics under different amplitude anisotropy conditions: (a)–(c) for $\gamma_a = 0$ and (d)–(f) for $\gamma_a = -30~\mathrm{GHz}$. (a), (d) Evolution of polarization states of a solitary VCSEL with increasing electrical pumping term ($\eta$). For $\gamma_a = 0$, the polarization switches from $x$-dominant (black curve) to $y$-dominant (red curve) as $\eta$ increases, while for $\gamma_a = -30~\mathrm{GHz}$, the polarization remains $x$-dominant. (b), (e) Both VCSELs are biased at $\eta = 1.1$, and the ML emits linearly polarized light at 45° injected into the SL at $t = 40~\mathrm{ns}$. (c), (f) Locking maps showing $\Delta \theta = |\theta_{\text{SL}} - \theta_{\text{ML}}|$ for varying $E_{\text{inj}}$ and frequency detuning ($\Delta \omega$) at $\eta = 1.1$.}
\end{figure*}

\begin{figure*}
\centering 
\includegraphics[height=0.3\textheight]{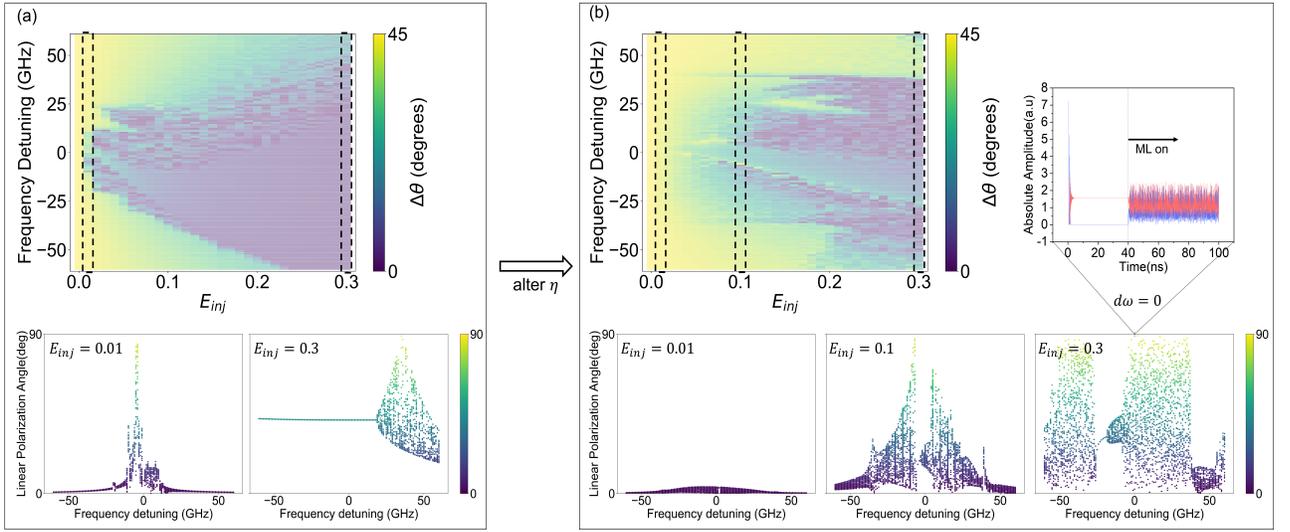} 
\caption{VCSEL OIL performance under different bias current conditions for $\gamma_a = 0$, with locking maps shown at $\eta = 1.2$ (a) and $\eta = 3.4$ (b). Below the locking maps, the linear polarization angles of the SLs under OIL are shown for certain $E_{\text{inj}}$ values, corresponding to the black-boxed regions in the locking maps. At $\eta = 3.4$, the polarization dynamic of the VCSEL under OIL with $E_{\text{inj}} = 0.3$ and $\Delta \omega = 0$ is illustrated in the top-right inset of (b).}
\end{figure*}

The parameters used in our SFM are summarized in Table 2, with values adjusted from our previous work to simulate an 850 nm VCSEL~\cite{Loke2023}. Here, $\kappa$ represents the cavity decay rate related to photon lifetime, while $\gamma$ represents the decay rate of the total carrier population. The parameter $\alpha$ is known as the linewidth enhancement factor. $\gamma_s$ accounts for the population differences between spin-up and spin-down states. $\eta$ represents the electrical pumping term. The gain anisotropy can be split into amplitude and phase terms, represented by $\gamma_a$ and $\gamma_p$, respectively. $\gamma_a$, influenced by factors such as anisotropic quantum well gain and cavity geometry, is the key parameter targeted for control in the first method, while $\gamma_p$ affects the resonance frequency. These two components are typically treated as independent factors~\cite{Martin1997}. The two linearly polarized modes have resonance frequencies given by $\omega_x = \alpha \gamma_a - \gamma_p$ and $\omega_y = \gamma_p - \alpha \gamma_a$, respectively. The magnitudes of $\Delta \phi$, $\Delta_x$, and $\Delta_y$ are given by $\Delta \phi = 2(\gamma_p - \alpha \gamma_a)t + \phi_y - \phi_x$, $\Delta_x = \omega_y t - \phi_x$, and $\Delta_y = \omega_x t - \phi_y$.

The ML output is set to emit linearly polarized light at $45^\circ$, with amplitude along the $x$- and $y$-directions set as $E_{\text{injx}} = E_{\text{injy}} = E_{\text{inj}}$, and the coupling coefficient is denoted as $K_{\text{inj}}$. In this simulation, the value of $E_{\text{inj}}$ varies from 0 to 0.3. The frequency detuning ($\Delta \omega$), defined as the difference between the ML and SL frequencies, is swept from $-60$\,GHz to $+60$\,GHz. For each value set of $E_{\text{inj}}$ and $\Delta \omega$, 100 simulations are conducted. For each value set, the average polarization difference ($\Delta \theta = \left|\theta_{\text{SL}} - \theta_{\text{ML}}\right|$) at $t = 100$\,ns is calculated. 

In the above experiments, modifying the arm length ratio of the cruciform or rotating its angle alters the amplitude anisotropy ($\gamma_a$) along two orthogonal directions.  For a solitary VCSEL with $\gamma_a = 0$ (Fig.~3(a)), polarization switching occurs with increasing injection current, whereas with strong $\gamma_a$ (Fig.~3(d)), the polarization state remains stable and does not switch. When the current is biased at the same level ($\eta = 1.1$, corresponding to Region~I) and the ML is injected, VCSELs with $\gamma_a = 0$ (Fig.~3(b)) achieve polarization locking, as indicated by the predominance of purple regions in the locking map (Fig.~3(c)). In contrast, for VCSELs with strong $\gamma_a$ (Fig.~3(e)), the polarization state remains unlocked, as indicated by the yellow regions in the locking map (Fig.~3(f)). When $\gamma_a$ is too strong, the gain in one direction far exceeds that in the orthogonal direction, allowing the inherent gain to dominate over the external photon injection.

Next, we set $\gamma_a = 0$ to isolate its impact and adjust the electrical pumping to $\eta = 1.2$ (the switching point between Region~II and Region~III) and $\eta = 3.4$ (Region~III). The corresponding locking maps are shown in Fig.~4(a) and Fig.~4(b), respectively. As $\eta$ increases from 1.2 to 3.4, the locking range becomes narrower, and the minimum $E_{\text{inj}}$ required to achieve locking increases. These changes suggest that higher bias currents make OIL more difficult, as the system requires stronger external injection to compensate for the reduced locking range.

For both current bias points, the locking range expands as $E_{\text{inj}}$ increases. Below the locking diagrams, the linear polarization angle of the SL under OIL is shown for specific values of $E_{\text{inj}}$, corresponding to the black-boxed regions in the locking maps, visualizing the results of all 100 runs for each frequency detuning value. For $\eta = 1.2$, two cases are highlighted: $E_{\text{inj}} = 0.01$ and $E_{\text{inj}} = 0.3$. At $E_{\text{inj}} = 0.01$, the SL is influenced by the ML only near zero frequency detuning. Increasing $E_{\text{inj}}$ to 0.3 extends the locking range,  the SL remains stably locked at $45^\circ$ for most $\Delta \omega$ values, with instability observed only for a small range of positive detuning. When $\eta$ is increased to 3.4, the SL initially remains unaffected by the ML at $E_{\text{inj}} = 0.01$. As $E_{\text{inj}}$ increases from 0.01 to 0.3, the locking range expands, with the polarization state of the SL being influenced by the ML across more frequency detuning values. The top-right inset in Fig.~4(b) zooms in on the case where $E_{\text{inj}} = 0.3$ and $\Delta \omega = 0$. Here, the SL's polarization state starts oscillating between two orthogonal states after the ML is turned on. This locking behavior is less stable compared to cases with lower $\eta$.

In summary, we first fabricate VCSELs with tailored aperture designs to control amplitude anisotropy. Then, we optimize the bias current to investigate its role in enhancing polarization locking.
We achieve a minimum injection power of $3.6~\mu\mathrm{W}$ and robust polarization locking across a frequency detuning range of $\pm 10~\mathrm{GHz}$.
The low minimum injection power and expanded locking range lower the difficulty of polarization locking, enhancing the potential for polarization encoding in optical computing applications, such as VCSEL-based Ising computers and optical neural networks~\cite{Utsunomiya2011,Loke2023,Babaeian2019,Gao2024,Chen2023,Lan2024,Zhang2025, Liu2025}. Moreover, in large-scale Ising computers, where numerous VCSELs share the same injection power, achieving polarization locking with the reduced injection power allocated to each VCSEL becomes a critical challenge~\cite{Zhang2025}, highlighting the importance of designing systems that facilitate polarization locking with ease. These improvements may further support emerging applications in computational imaging~\cite{chen2024dual,lin2025geocomplete} and machine learning~\cite{lin2025rgb, Chen2026hypo, chen2025auto, Teng2025, Du2026, lin2025nighthaze, lin2024nightrain}.

\begin{acknowledgments}
This work was supported by the National Research Foundation, Singapore, under its Competitive Research Programme (NRF CRP24-2020-0003) and by both the National Research Foundation, Singapore, and A*STAR under the Quantum Engineering Programme (NRF 2021-QEP2-02-P12).
\end{acknowledgments}

\section*{Conflict of interest}
The authors have no conflicts to disclose.

\section*{Data Availability}

The data that support the findings of this study are available
from the corresponding author upon reasonable request.

\section*{References}
\nocite{*}
\bibliographystyle{unsrt}
\bibliography{references}

\end{document}